\documentclass[12pt]{article}
\usepackage{hyperref}
\usepackage[pdftex]{graphicx}
\usepackage{amsmath,amssymb}
\usepackage{cancel}
\usepackage{appendix}
\newcommand{\gs}{\ensuremath{g_s}} 
\newcommand{\ls}{\ensuremath{l_s}} 

\def\eps{\ensuremath{\epsilon}}

\def\p{\partial}

\newcommand{\cN}{{\mathcal{N}}}
\newcommand{\cO}{{\mathcal{O}}}

\newcommand{\Xp}{{X^{'}}}
\newcommand{\Xd}{{\dot{X}}}
\newcommand{\vX}{\mbox{$\vec{X}$}}

\newcommand{\be}{\begin{equation}}
\newcommand{\ee}{\end{equation}}
\newcommand{\bea}{\begin{eqnarray}}
\newcommand{\eea}{\end{eqnarray}}

\begin{document}

\begin{titlepage}

\begin{center} \Large \bf No Line on the Horizon: On Uniform Acceleration \\
 and Gluonic Fields at Strong Coupling
\end{center}

\begin{center}
J.~Antonio Garc\'{\i}a\footnote{garcia@nucleares.unam.mx},
Alberto G\"uijosa\footnote{alberto@nucleares.unam.mx}
and Eric J.~Pulido\footnote{ericpulido@gmail.com}

\vspace{0.2cm}
Departamento de F\'{\i}sica de Altas Energ\'{\i}as, Instituto de Ciencias Nucleares, \\
Universidad Nacional Aut\'onoma de M\'exico,
\\ Apartado Postal 70-543, M\'exico D.F. 04510, M\'exico\\
\vspace{0.2cm}
\end{center}

\begin{center}
{\bf Abstract}
\end{center}
\noindent We study a few assorted questions about the behavior of strings on anti-de Sitter spacetime (AdS), or equivalently, `flux tubes' in strongly-coupled conformal field theories (CFTs). For the case where the `flux tube' is sourced by a uniformly accelerated quark (or, more generally, a quark that asymptotes to uniform acceleration in the remote past), we point out that the dual string embedding known heretofore terminates unphysically at the worldsheet horizon, and identify the correct continuation, which is found to encode a gluonic shock wave. For arbitrary quark motion, we  show that, contrary to common understanding, the worldsheet horizon does \emph{not} in general represent a dividing line between the portions of the string respectively dual to the quark and to the gluonic radiation emitted by it.

\vspace{0.2in}
\smallskip
\end{titlepage}
\setcounter{footnote}{0}

\tableofcontents

\section{Introduction and Summary}\label{introsec}

One of the many interesting entries in the dictionary of gauge/gravity dualities \cite{malda,gkpw} identifies flux tubes in strongly-coupled gauge theories with strings in the corresponding gravity theories (see, e.g., \cite{brandhuber,gross,barflux,klebanov}). This connection of course harks back to the much studied QCD string \cite{thooft,rebbi,nambu,luscher}, but in the gauge/gravity context, there are a few surprises: the string is infinitesimally thin, it lives in higher dimensions, and it exists even in nonconfining setups such as conformal field theories (CFTs)! Of course, in the latter case it is not literally a flux \emph{tube}, but tends to spread Coulombically \cite{maldawilson,dkk} (so perhaps it would be more fitting to call it a flux spray). The AdS/CFT correspondence allows us to study the time-dependent spread of the gluonic field in a CFT at strong coupling by examining the dynamics of a string that moves on anti-de Sitter (AdS) spacetime. This has been done in many works, beginning with \cite{dkk,cg} and continuing on to the recent contributions \cite{liusynchrotron,veronika,iancu1,iancu2,trfsq,iancu3} (for similar computations at finite temperature, see, e.g., \cite{gluonicprofile,rajagopalshining}).

In more detail, a string on AdS that extends all the way to the boundary is dual to an infinitely heavy quark in the CFT, with the endpoint of the string located on the AdS boundary corresponding to the quark (understood as a bare source in the fundamental representation of the gauge group) and the body of the string codifying the near and radiation fields sourced by the quark. The string embedding causally associated with \emph{any} timelike trajectory of the quark was worked out in \cite{mikhailov}. When the quark moves at constant velocity, the string translates uniformly, extending purely along the radial AdS direction, and is of course the embodiment of a boosted Coulomb profile. When the quark accelerates, the string trails behind its endpoint, and sucks energy out of it, expressing the fact that gluonic radiation is emitted in the CFT \cite{lorentzdirac,damping}. In this case, a black hole appears on the string worldsheet \cite{dragtime}, which signals energy loss just
  as in the case of a quark moving through a thermal plasma \cite{gubserqhat,ctqhat} (dual to a string on the Schwarzschild-AdS geometry).

 In the present paper, we will examine a few assorted questions in this context. After setting the stage in Section \ref{mikhailovsec}, we show in Section \ref{whippingsec} that, in spite of the fact that the string has a general tendency to fall down (i.e., away from the boundary) in the AdS geometry, there exist endpoint motions for which the string actually turns up for some distance before bending back down. Through the UV/IR connection \cite{uvir}, this can be interpreted in CFT language as the statement that, in spite of the fact that the flux tube sourced by the quark has a general tendency to spread, there exist quark motions for which the flux tube actually refocuses for some distance before starting to spread again.

 A particularly peculiar example is the case where the quark undergoes uniform acceleration \cite{xiao,ppz,brownian}. In this case the string takes the shape of the arc of a circle, the worldsheet black hole is static, and, as we show in Section \ref{accelsec} (see also \cite{veronikasemenoff}), the string profile of \cite{mikhailov} actually \emph{terminates} at the location of the worldsheet horizon! This is not a mistake: this much string really codifies the gluonic fields generated by the uniformly accelerating quark since the beginning of time. But of course, the string cannot just end in midair.

 Interestingly, the only smooth extension is to continue the circular shape all the way back to the boundary, thereby obtaining a configuration dual to a  quark \emph{and antiquark} that accelerate uniformly back to back \cite{xiao}. If this were indeed the only allowed continuation, then the AdS/CFT correspondence would be making the prediction that it is impossible to have an isolated quark undergoing uniform acceleration, as if this type of motion were so violent that the flux tube emerging from the quark is forced to refocus all the way down to zero width, onto an antiquark. We find, however, that there does exist a string embedding corresponding to an isolated quark. The segment of string beyond the worldsheet horizon is purely radial and moves at the speed of light. It corresponds to a gluonic shock wave that exists already at the beginning of time and is shed by the quark, because the perturbation in the gluonic field produced by the quark's deceleration cannot
 catch up with the shock wave. {}From the analysis it becomes clear that this story holds not only for uniform acceleration, but for any quark trajectory that asymptotically approaches uniformly accelerated motion in the remote past.

It should be borne in mind that the abrupt termination of the dual string at the horizon is a phenomenon specific to the uniformly accelerated quark: even though a horizon does appear on the worldsheet whenever the quark/endpoint accelerates in any fashion \cite{dragtime}, in general the string embedding obtained in \cite{mikhailov}  \emph{does not} terminate there. In other words, the gluonic fields generated by the quark since the beginning of time are generally codified not only by the string segment outside the worldsheet horizon, but also by the segment inside.

The fact that the horizon divides the string dual to an accelerating quark into 2 segments is reminiscent of the fact that the total energy-momentum of the string is the aggregate of 2 distinct contributions, corresponding to the intrinsic and radiated energy-momentum of the quark, which have been separated via a worldsheet analysis in \cite{mikhailov,dragtime,lorentzdirac,damping} (and at the spacetime level in \cite{tmunu}).  It is natural to conjecture, then, that the worldsheet horizon  should be physically interpreted as a dividing line between the portions of the string that are respectively dual to the near and radiation fields \cite{dragtime,dominguez,xiao,beuf,veronikamukund,iancu2,rajagopalshining}.
 This would imply in particular that the rate of energy-momentum flow across the horizon represents the rate at which the quark radiates energy-momentum. The latter is known from \cite{mikhailov} to be proportional to the familiar Lienard formula from classical electrodynamics\footnote{This is true for an infinitely heavy quark; the corresponding formula for finite quark mass was worked out in \cite{dragtime,lorentzdirac,damping}.}. In consonance with the interpretation of the horizon as a dividing line, in \cite{xiao} it was shown that in the case of uniform acceleration, the rate of energy flow across the horizon does in fact correctly reproduce the Lienard formula.

 In Section \ref{flowsec} we pursue this same issue for arbitrary
  quark motion, by calculating the rate of energy-momentum flow across an arbitrary curve on the worldsheet. We discover that the agreement found in \cite{xiao} is due to the special features of the trajectory considered there, and in general, the flow of energy-momentum across the worldsheet horizon does \emph{not} reproduce the Lienard formula. More generally, we show that the information on the near and radiation fields of the quark is \emph{not} separated geometrically on the worldsheet at fixed observation time $t$: it is in general spread throughout the entire string, and can only be explicitly disentangled through the algebraic-differential procedure of \cite{mikhailov,dragtime,tmunu}. Curiously, if we consider instead the string at a fixed value of the retarded time $t_r$ involved in the construction of \cite{mikhailov} (see (\ref{mikhsolnoncovariant}) below), then a geometric separation can in fact be made, but the quark is found to correspond to the \emph{entire} string, and all of the radiation lies at the Poincar\'e horizon of AdS.

\section{Quarks, with Strings Attached}\label{mikhailovsec}

Our discussion applies to any instance of the AdS$_{d+1}$/CFT$_{d}$ correspondence, but for concreteness we will present it in terms of the best understood example, which identifies the maximally supersymmetric (i.e., $\cN=4$) $SU(N_c)$ Yang-Mills theory (MSYM) on $3+1$-dimensional Minkowski spacetime with Type IIB string theory on an asymptotically AdS$_5\times S^5$ geometry (sustained by $N_c$ units of flux of the Ramond-Ramond 5-form field strength). A heavy quark propagating in the symmetric vacuum of MSYM is described in dual language by a string moving on the Poincar\'e wedge of the AdS$_5$ geometry,
\begin{equation}\label{metric}
ds^2=G_{mn}dx^m dx^n={R^2\over z^2}\left(
-dt^2+d\vec{x}^{\,2}+dz^2 \right)
\end{equation}
(and lying at a constant position on the $S^5$, consistent with the equations of motion).
The coordinates $x^{\mu}\equiv(t,\vec{x})$  parallel to the AdS boundary $z=0$  are
 identified with the gauge theory spacetime coordinates, whereas the
radial direction $z$ is mapped to a variable length (or, equivalently, inverse energy) scale in MSYM
\cite{uvir}.  The MSYM coupling is connected to the string coupling via $g_{YM}^2=4\pi\gs$. The radius of curvature $R$ is related to the gauge theory 't Hooft coupling $\lambda\equiv g_{YM}^2 N_c$ through
\begin{equation}\label{lambda}
 \lambda={R^4\over \ls^4}~,
\end{equation}
 where $\ls$ denotes the string length. Throughout this paper we will consider for simplicity the case of an infinitely massive quark, which corresponds to a string extending all the way from the Poincar\'e horizon at $z\to\infty$ to the AdS boundary at $z=0$. Following \cite{dragtime,lorentzdirac,damping}, it would be easy to generalize our analysis to the case where the quark is heavy but has finite mass $m$, where the string would end on $N_f$ flavor D7-branes that extend from the boundary down to  a radial location $z_m$ such that \cite{kk}
 \begin{equation}\label{zm}
 m=\frac{\sqrt{\lambda}}{2\pi z_m}~.
 \end{equation}

In the large $N_c$, large $\lambda$ limit, the string embedding is determined by extremizing the Nambu-Goto action
\begin{equation}\label{nambugoto}
S_{\mbox{\scriptsize NG}}=-{1\over 2\pi\ls^2}\int
d^2\sigma\,\sqrt{-\det{g_{ab}}}
=-{1\over 2\pi\ls^2}\int
d^2\sigma\,\sqrt{\left(\Xd\cdot\Xp\right)^2-\Xd^2\Xp^2}~,
\end{equation}
where $g_{ab}\equiv\p_a X^m\p_b X^n G_{mn}(X)$ ($a,b=0,1$) is
the induced metric on the worldsheet, and of course $\,\dot{}\equiv\p_{\sigma^{0}}\equiv\p_{\tau}$, ${}^{\prime}\equiv\p_{\sigma^1}\equiv\p_{\sigma}$.
For the most part we will work in the static gauge
$\tau=t$, $\sigma=z$, where the string embedding is described as $\vX(t,z)$.

The quark's trajectory coincides with the path of the string endpoint at the AdS boundary,
\begin{equation}\label{quarkx}
x^{\mu}(\tau)=X^{\mu}(\tau,\sigma)|_{z=0}~.
\end{equation}
For any timelike quark trajectory $x^{\mu}(\tau_r)$, parametrized by \emph{proper} time $\tau_r$, the dual string embedding that corresponds to a \emph{purely outgoing} gluonic field configuration was found by Mikhailov \cite{mikhailov},
\begin{equation}\label{mikhsol}
 X^{\mu}(\tau_r,z)=x^{\mu}(\tau_r)+zv^{\mu}(\tau_r)~,
\end{equation}
where $v^{\mu}\equiv dx^{\mu}/d\tau_r$ denotes the 4-velocity of the quark (such that $\eta_{\mu\nu}v^{\mu}v^{\nu}=-1$).
Combining (\ref{metric}) and (\ref{mikhsol}), the induced metric on the worldsheet is determined to be
\begin{equation}\label{wsmetric}
g_{\tau_r\tau_r}={R^2\over
z^2}(z^2 a^2-1),\qquad g_{zz}=0,\qquad g_{z\tau_r}=-{R^2\over z^2},
\end{equation}
where of course $a^{\mu}\equiv dv^{\mu}/d\tau_r$. Notice that this
implies in particular that the constant-$\tau_r$ lines are null.

In non-covariant notation, (\ref{mikhsol}) dictates that
\begin{eqnarray}\label{mikhsolnoncovariant}
t(t_r,z)&=&t_r+\frac{z}{\sqrt{1-\vec{v}(t_r)^2}}~,\\
\vec{X}(t_r,z)&=&\vec{x}(t_r)+\frac{\vec{v}(t_r) z}{\sqrt{1-\vec{v}(t_r)^2}}~, \nonumber
\end{eqnarray}
where we have used $d\tau_r=dt_r/\gamma$, $v^{\mu}=\gamma(1,\vec{v})$.
{}From (\ref{mikhsolnoncovariant}) we learn that the behavior of the string at a given time $t=X^0$ and radial depth $z$ (which essentially encodes the gluonic field a distance $z$ away from the quark) is parametrized in terms of the behavior of the quark/string endpoint at the earlier, \emph{retarded} time $t_r$, in complete analogy with the Lienard-Wiechert story in classical electrodynamics. We also see that the null curves at fixed $t_r$ (or $\tau_r$) trace out straight lines in the AdS coordinates (\ref{metric}), implying that the embedding (\ref{mikhsolnoncovariant}) is a ruled surface. We will refer to these as Mikhailov lines; their importance stems from the fact that endpoint information propagates along them.

It follows from (\ref{mikhsolnoncovariant}) that
\begin{eqnarray} \label{Xdot}
\dot{\vec{X}}&\equiv&
\left(\frac{\partial \vec{X}}{\partial t}\right)_{z}
=\vec{v}+\frac{(1-\vec{v}^{\,2})\vec{a}z}{(\vec{v}\cdot\vec{a})z
+(1-\vec{v}^{\,2})^{3/2}}~,
\\
\label{Xprime}
\vec{X}^{'}&\equiv&\left(\frac{\partial\vec{X}}{\partial z}\right)_{t}
=-\frac{\sqrt{1-\vec{v}^{\,2}}\,\vec{a}z}{(\vec{v}\cdot\vec{a})z
+(1-\vec{v}^{\,2})^{3/2}}~.
\end{eqnarray}
Here and in what follows, the quark/endpoint velocity $\vec{v}$ and acceleration $\vec{a}$  are understood to be evaluated at the appropriate $t_r(t,z)$. Using (\ref{Xdot}) and (\ref{Xprime}), it was shown in \cite{mikhailov} that
the total energy of the string embedding (\ref{mikhsol}) at some observation time $t$ (which is conserved in the $N_c\to\infty$ limit) can be reexpressed (via a
change of integration variable $z\to t_r$) as a local functional of the quark trajectory,
\begin{equation}\label{emikh}
E(t)={\sqrt{\lambda}\over 2\pi}\int^t_{-\infty}dt_r
\frac{\vec{a}^{\,2}-\left[\vec{v}\times\vec{a}\right]^2}{\left(1-\vec{v}^{\,2}\right)^3}
+E_q(\vec{v}(t))~.
\end{equation}
The second term in the above equation arises
from a total derivative that was not explicitly written down in \cite{mikhailov}, but was shown in \cite{dragtime} to take the form
\begin{equation}\label{edr}
E_q(\vec{v})={\sqrt{\lambda}\over
2\pi}\left.\left({1\over\sqrt{1-\vec{v}^{\,2}}}{1\over
z}\right)\right|^{z_m=0}_{\infty}=\gamma m~,
\end{equation}
thus giving the expected Lorentz-invariant dispersion relation for
the quark. The energy split achieved in (\ref{emikh}) therefore
admits a transparent and pleasant physical interpretation: $E_q$ (associated only
with information of the string endpoint) is the
intrinsic energy of the quark at time $t$, and the integral over
$t_r$ (associated with the body of the string) encodes the accumulated energy \emph{lost} by the quark to its gluonic field
over all times prior to $t$. Completely analogous statements can be derived for the spatial momentum. In covariant notation, the rate of energy-momentum radiated by the quark (incorporating the integrand in the first term of (\ref{emikh})) is given by
\begin{equation}\label{lienard}
\frac{dp^{\mu}_{ \scriptstyle \mathrm{rad}}}{d\tau_r}=\frac{\sqrt{\lambda}}{2\pi}a^2 v^{\mu}~.
\end{equation}
Remarkably,  aside from the overall normalization, this is in precise agreement with the standard Lienard formula from classical electrodynamics.
The AdS/CFT correspondence thus teaches us that, in this unfamiliar quantum nonlinear setting, the energy-momentum loss turns out to depend only locally on the quark worldline. The way in which the radiation rate and dispersion relation are modified when the quark has finite mass was established in \cite{dragtime,lorentzdirac,damping}.

\section{Whipping Strings} \label{whippingsec}

We see from (\ref{mikhsolnoncovariant}) that, when we consider the profile of the string at a given observation time $t$, we generically have $z\to\infty$ as $t_r\to -\infty$, so points at earlier and earlier retarded times are forced to lie at larger and larger values of $z$. This is the reason why the string extends all the way from the AdS boundary ($z=0$) to the Poincar\'e horizon ($z\to\infty$), as expected for an isolated quark. The entire string is needed to codify the gluonic field emitted by the quark since the beginning of time.

For generic trajectories, $z$ will increase monotonically as we move along the string starting from the endpoint at the AdS boundary. This is consistent with the fact that the string has a tendency to fall down towards larger $z$, as a result of the AdS curvature. In gauge theory language, this just says that the flux tube has a general tendency to spread.

It is interesting to ask whether this spread can stop momentarily, or perhaps even reverse for some distance. A naive use of the UV-IR connection \cite{uvir} translates this into the question of whether the string can become horizontal (i.e., parallel to the boundary) at some point, or perhaps even have a segment that approaches, rather than recedes from, the AdS boundary. In actual fact, in dynamical situations the mapping between the string and flux tube profiles can be more subtle, and depends on the observable we employ to establish it. If we consider only the expectation value of local operators, such as the energy-momentum tensor or the Lagrangian density, then an important role is played by the propagation time back to the boundary of the corresponding bulk field sourced by the string segment in question. Explicit calculations show that the contributions of the various points along the string embedding (\ref{mikhsol}) take the form of a total derivative, and so the net
  result depends only on the position of the quark/endpoint at a particular retarded time \cite{iancu2,beaming}. If, on the other hand, we probe the string  at points away from the boundary by considering nonlocal quantities such as Wilson loops or higher-point correlators \cite{precursors}, then its instantaneous profile can be interpreted in accord with the naive UV-IR connection. It is this perspective that we will adopt in the following discussion.

The first question we wish to ask is whether the string can become horizontal at a point. The easiest way to answer this is to read off from (\ref{mikhsolnoncovariant}) that $z(t,t_r)=\sqrt{1-\vec{v}^{\,2}}(t-t_r)$, and extremize by setting $(\p z/\p t_r)_t=0$. A solution can be found only if $\vec{v}\cdot\vec{a}<0$. Given any point $t_r$ on the quark/endpoint trajectory where this condition is satisfied (i.e., where the acceleration opposes the motion), the string will become horizontal at a point along the corresponding Mikhailov line that lies at the later time
\begin{equation}\label{textremum}
t=t_r-\frac{1-\vec{v}^{\,2}}{\vec{v}\cdot\vec{a}}~,
\end{equation}
or equivalently, at the radial depth
\begin{equation}\label{zextremum}
z=-\frac{(1-\vec{v}^{\,2})^{3/2}}{\vec{v}\cdot\vec{a}}~.
\end{equation}
(One can check in (\ref{Xprime}) that indeed $\vec{X}^{'}\to\infty$ at this point.) This shows that actually it is not all that rare for the flux tube to momentarily stop spreading.

 The next question concerns the nature of this extremum. By computing $(\p^2 z/\p t_r^2)_t$, it is easy to establish that the point (\ref{zextremum}) on the string  is a maximum if the jerk $\vec{j}\equiv d\vec{a}/dt_r$ at $t_r$ satisfies
\begin{equation}\label{jerkineq}
\vec{j}\cdot\vec{v}+\vec{a}^{\,2}+3\frac{(\vec{v}\cdot\vec{a})^2}{1-\vec{v}^{\,2}}
=\vec{j}\cdot\vec{v}+\vec{a}^{\,2}+3\frac{(1-\vec{v}^{\,2})^{2}}{z^2}>0~.
\end{equation}
When (\ref{jerkineq}) holds, the string turns back towards the AdS boundary, or equivalently, the MSYM flux tube refocuses. When the string profile at a given observation time $t$ does turn around in this manner, it generically turns back away from the boundary at some earlier value of $t_r$, for which the reversed version of the inequality (\ref{jerkineq}) is satisfied. This can only happen if $\vec{j}\cdot\vec{v}<0$, and becomes more and more difficult as $z\to 0$, because the required $|\vec{j}|\to\infty$. This means that, if we want the string to whip the string to make it return arbitrarily close to the AdS boundary, we must be prepared to jerk the endpoint arbitrarily violently.

Our overall conclusion, then, is that there do exist not terribly contrived quark/end-point trajectories which whip the string in such a way that its profile no longer increases monotonically towards $z\to\infty$. Under these circumstances, the dual flux tube (when examined with an appropriate non-local probe, as discussed above) reverses its general tendency to spread, and in fact refocuses for some distance, before starting to spread again.

\section{Uniform Acceleration}\label{accelsec}

A simple case that illustrates the whipping behavior of the previous section is
a uniformly accelerated quark with proper acceleration $A$,
\begin{equation}\label{quarkaccel}
x(t_r)=\sqrt{A^{-2}+t_r^2}~.
\end{equation}
The proper time of the quark is
\begin{equation}\label{tau}
\tau_r=A^{-1}\mathrm{arcsinh}(At_r)~,
\end{equation}
and the general solution (\ref{mikhsolnoncovariant}) takes the form
\begin{eqnarray}\label{mikhaccel}
t(t_r,z)&=&t_r+\sqrt{1 +A^{2}t_r^2}\,z~\\
X(t_r,z)&=&\sqrt{A^{-2}+t_r^2}+At_r z~.\nonumber
\end{eqnarray}
Using (\ref{zextremum}) and (\ref{jerkineq}), it is easy to check that, for all $t>0$, the quark worldline (\ref{quarkaccel}) leads to a turn-around point (a maximum in $z(t_r)$) in the string profile, located at $z=\sqrt{A^{-2}+t^2}$, $X=0$.

Upon eliminating $t_r$, (\ref{mikhaccel}) becomes \cite{brownian}
\begin{equation}\label{xiaoaccel}
X(t,z)=\pm\sqrt{A^{-2}+t^2-z^2}~.
\end{equation}
 This solution was found independently in \cite{xiao,ppz}. Knowing that  (\ref{mikhaccel}) is a special case of (\ref{mikhsolnoncovariant}), we are assured that it is the retarded embedding that codifies the physics of interest to us. The resulting worldsheet metric (\ref{wsmetric}) can easily be seen to have an event horizon at $z_h\equiv A^{-1}$ \cite{xiao,brownian}.

 A very peculiar property of (\ref{mikhaccel}) is that the coefficient of $z$ in $t(t_r,z)$ diverges linearly as $t_r\to-\infty$. This negates the generic conclusion we reached in the first paragraph of the previous section: due to the rate at which the quark/endpoint asymptotically approaches the speed of light, the string in this case \emph{does not} extend all the way down to $z\to\infty$. As $t_r\to -\infty$, we have $t\simeq t_r + A|t_r|z$, and thus $z\to z_h$. This means that the string only spans an arc of the circle (\ref{xiaoaccel}), extending from the boundary to the worldsheet horizon.

\begin{figure}[htb]
\centering
\includegraphics[width=12cm]{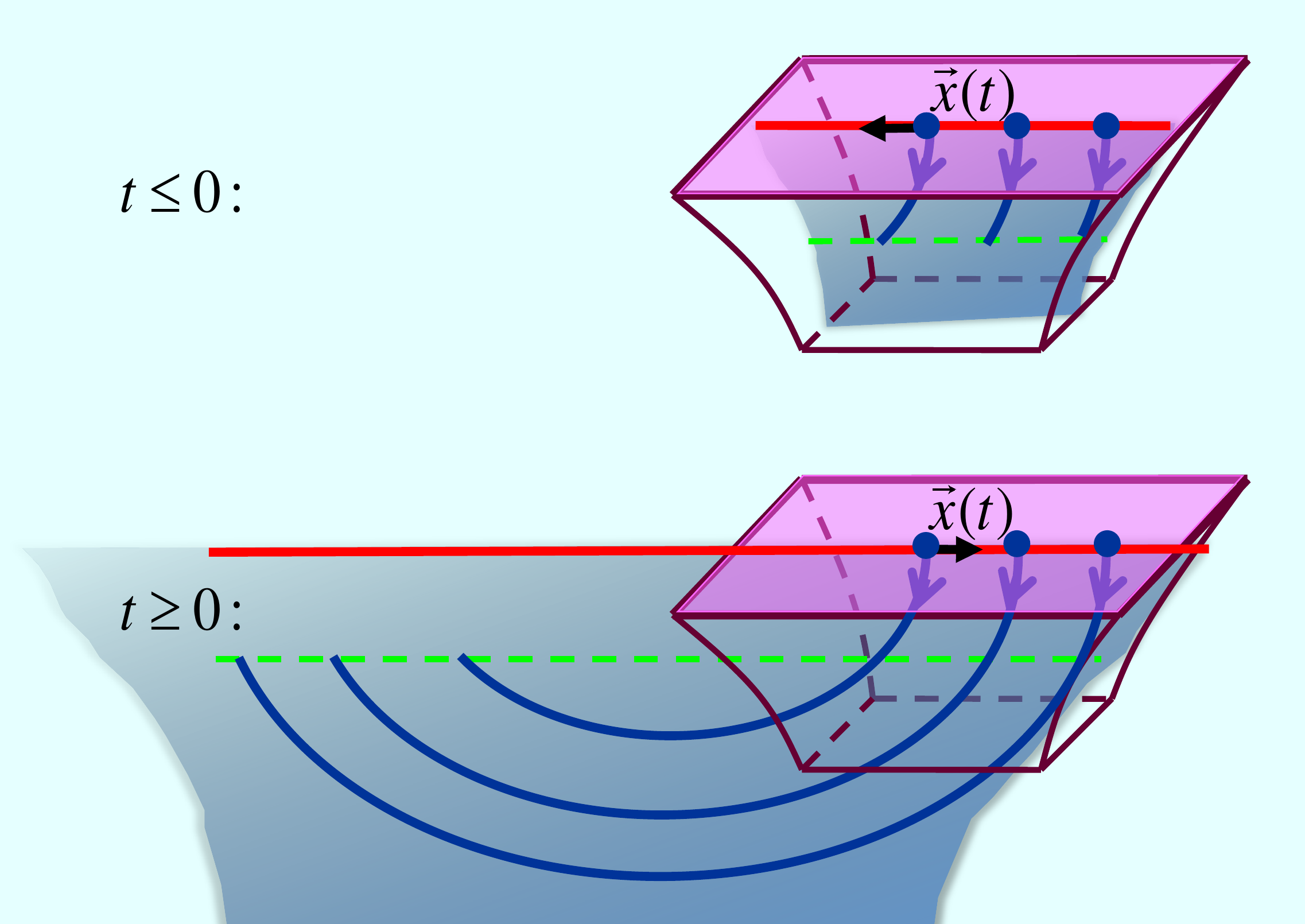}
\caption{Schematic representation of successive snapshots of the string dual to a uniformly accelerated quark, at various negative times (top) and positive times (bottom), showing that at any given instant, the Mikhailov embedding (\ref{mikhaccel}) dual to a quark undergoing uniform acceleration traces an arc that terminates at the worldsheet horizon (indicated by the green dashed line).
}\label{arcsfig}
\end{figure}

\begin{figure}[htb]
\centering
\begin{minipage}[t]{7.5cm}
\includegraphics*[trim= 10 60 10 75,width=5.6cm]{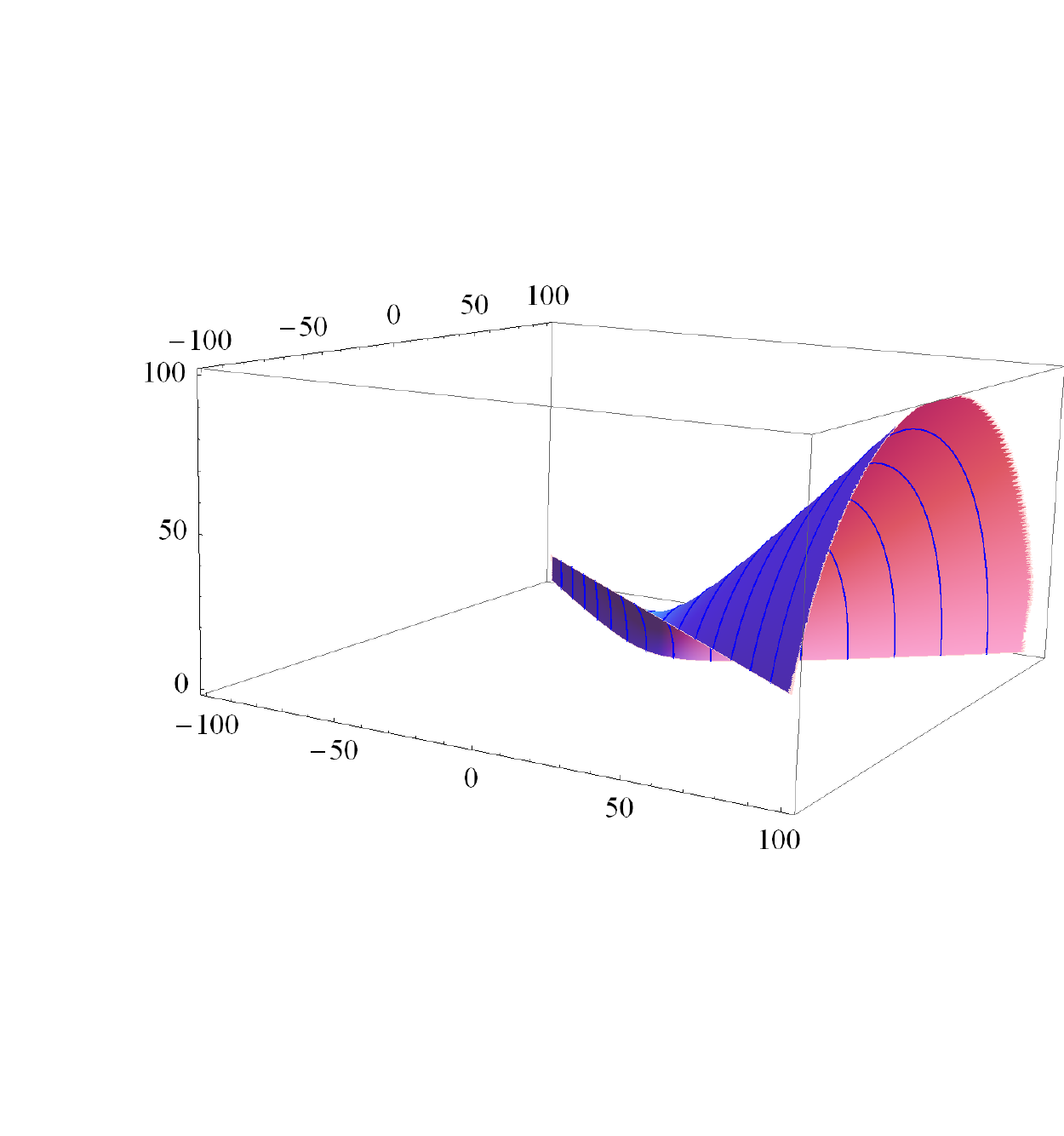}
\end{minipage}
\begin{minipage}[t]{7.5cm}
\includegraphics*[trim= 10 60 10 75,width=5.6cm]{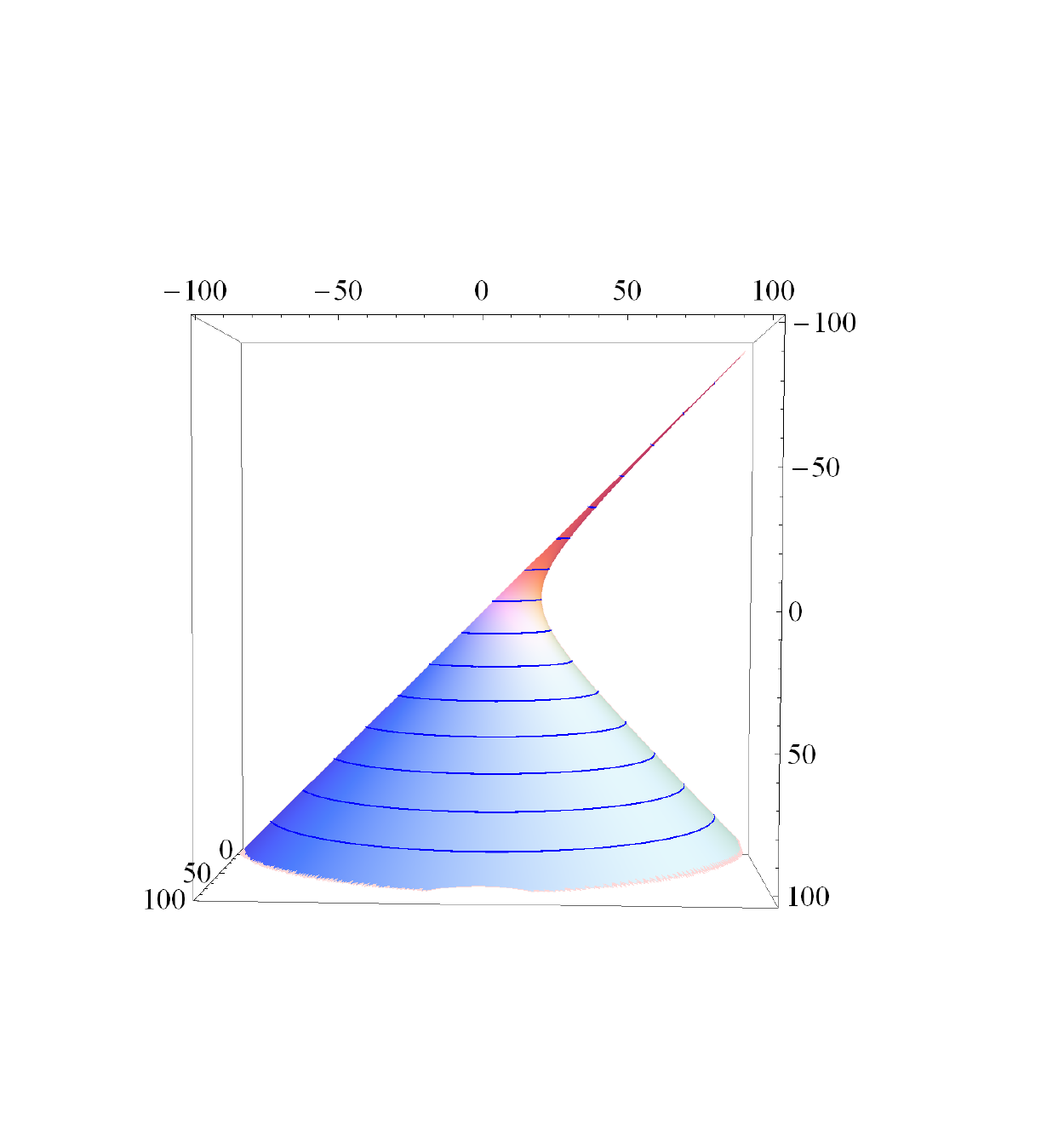}
\end{minipage}
\vspace*{-0.8cm}
\caption{Side and top views of the worldsheet (\ref{mikhaccel}) dual to a uniformly accelerating quark (in units such that $A=0.05$), plotted with Mathematica. The surface terminates abruptly at the location of the horizon ($z_h=20$, in these units). The (blue) curves indicate string contours at fixed $t$, and the $z$ axis points upward.
}
\label{uniformfig}
\end{figure}

 Moreover, unlike what (\ref{xiaoaccel}) by itself would have suggested, the string embedding (\ref{mikhaccel}) is \emph{not} symmetric under $t\to -t$. The point at $t_r\to -\infty$ ($z=A^{-1}$)  follows the trajectory $X=-t$, i.e., it always moves at the speed of light, in the $-x$ direction. This implies that, as seen in Figs.~\ref{arcsfig} and~\ref{uniformfig}, for $t<0$ the string terminates at the point where it first touches the horizon, and therefore covers less than a quarter-circle; whereas for $t>0$, it crosses the horizon, turns around at $z_{\mbox{\scriptsize max}}(t)\equiv \sqrt{A^{-2}+t^2}$,
 and then ends at the second point of intersection with the horizon, thus covering more than a quarter-circle.

 It is important to note that this is not a mistake: by construction, the embedding (\ref{mikhsolnoncovariant}) represents the entire portion of string influenced by the endpoint motion since $t_r\to -\infty$, and therefore codifies the entire contribution to the gluonic field produced by the quark since the beginning of time. This property can be understood as an extreme case of the beaming effect discovered in \cite{beaming}. It was shown in that work that, at any given instant when the quark/endpoint velocity is relativistic, the corresponding Mikhailov line remains close to the AdS boundary for a long time, implying that the contribution to the gluonic field encoded by it is collimated. In the case at hand, this beaming effect is maximal for $t_r\to-\infty$, and the Mikhailov line in question  is still close to the AdS boundary even after an \emph{infinite} amount of time has elapsed! (This line is in fact parallel to the boundary, and coincides with the horizon $z=A^{-1}$.) This is why, for this type of motion, $t_r\to-\infty$ does not translate into $z\to\infty$.

 Still, it is evident that the string cannot just end in midair, so an appropriate continuation must exist.
If we demand that the continuation be \emph{smooth}, then it is unique: the string must continue along the full semicircle (\ref{mikhaccel}), all the way back to the boundary. This embedding describes a quark \emph{and antiquark} that undergo uniform acceleration back to back (it was precisely in this guise that (\ref{xiaoaccel}) was found in \cite{xiao}).\footnote{The arc of string that is attached beyond the termination point of (\ref{mikhaccel}) corresponds to a \emph{purely ingoing} version of the string/gluonic wave of \cite{mikhailov}, obtained by inverting the signs in the terms linear in $z$. This explains the asymmetry between the portions of string respectively associated with the quark and the antiquark.}

If this were indeed the only possibility, then the AdS/CFT correspondence would be making the  prediction that it is impossible to have an isolated quark undergoing uniform acceleration, as if this type of motion were so violent that it forces the associated flux tube to refocus all the way down to zero width, thereby implying the existence of an appropriate color sink, an antiquark. This potential prediction, however, is surprising to the point that it is difficult to believe  it. {}From the gauge theory perspective, the absence of confinement leaves us with no reason whatsoever to suspect that a quark would necessarily have to be accompanied by an antiquark, and just for this particular type of motion.

\begin{figure}[htb]
\centering
\includegraphics[width=12cm]{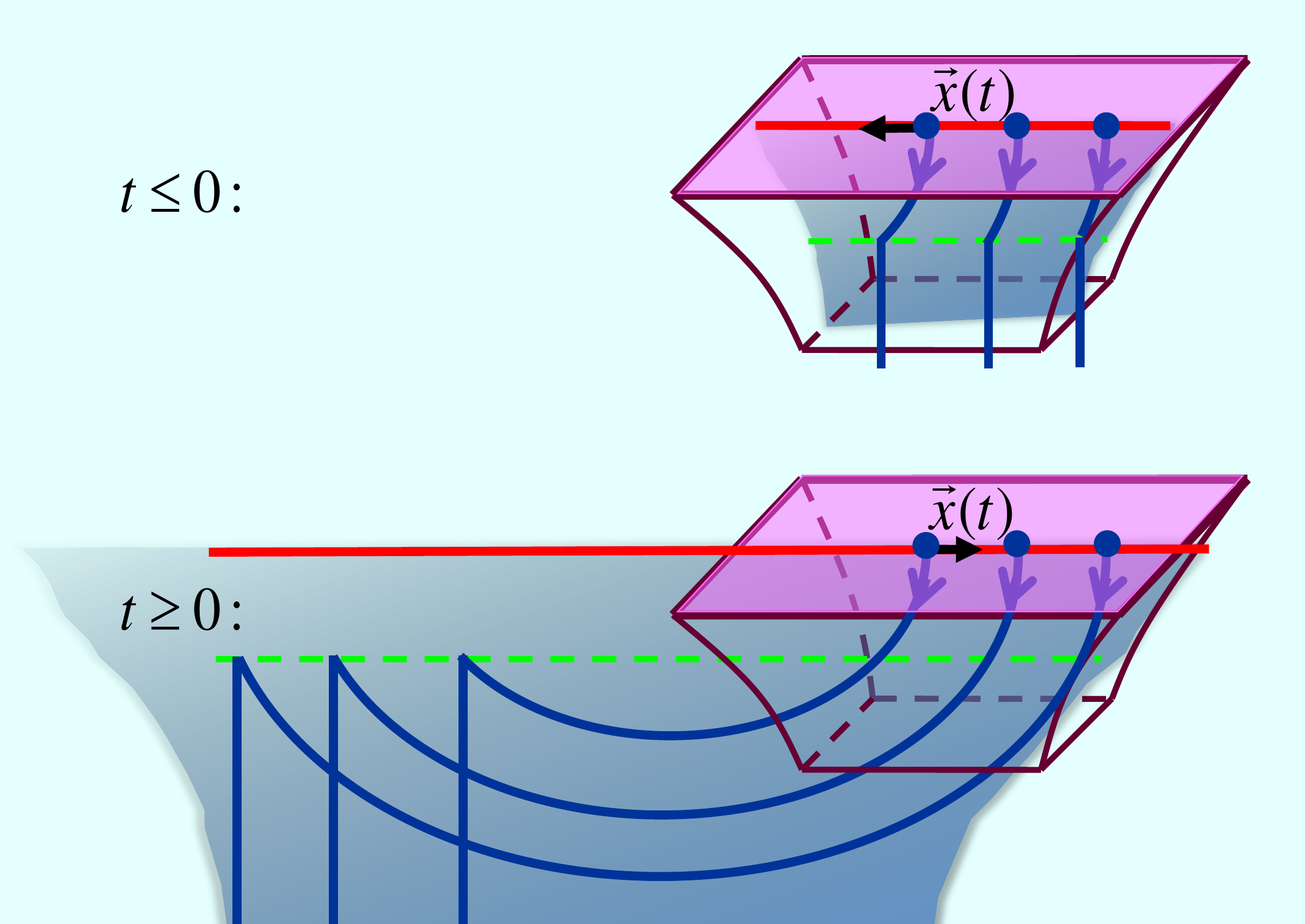}
 \caption{Successive snapshots of the \emph{complete} string embedding dual to a uniformly accelerated quark, at various negative times (top) and positive times (bottom). At any given instant, the arc (\ref{mikhaccel}) is continued by a vertical segment extending downward from the horizon, and moving to the left at the speed of light.
}
\label{arcscuspfig}
\end{figure}

 We are thus led to search for a continuation of the string that does not return to the AdS boundary, even if this entails a discontinuous slope at the juncture point, $z=A^{-1}$. In this search, we are guided by 2 pieces of information. First, at $t\to-\infty$ the known portion of the string is vertical (i.e., purely radial) and moves at the speed of light in the $-x$ direction. Second, for all $t$, the juncture point moves at the speed of light in that same direction. It is thus natural to conjecture that the missing portion of string is always vertical and moves at constant velocity. This would mean in particular that, at $t\to-\infty$, the entire string starts out vertical, correctly matching onto our desired initial condition: an isolated quark moving at constant velocity $v=-1$. As the quark/endpoint begins to slow down, a kink develops at $z=A^{-1}$, which becomes more and more pronounced as time progresses, eventually evolving into a cusp. The entire evolution can be
 seen in Fig.~\ref{arcscuspfig}.

 To put this conjecture to the test, we can argue as follows. We have seen above that the reason why the string (\ref{mikhaccel}) is incomplete is the fact that the endpoint asymptotically approaches the speed of light. So, if we deform the quark/endpoint trajectory such that this asymptotic behavior is modified, we should obtain a string that correctly extends all the way to $z\to\infty$. Upon removing this deformation to return to the case of uniform acceleration, we can then follow the fate of the additional portion of the string, and see if it conforms to our expectations.

 The first step then is to add to (\ref{quarkaccel}) some perturbation $\delta\vec{ x}(t_r)$ such that at $t_r\to-\infty$ the quark/endpoint now approaches a  subluminal speed, or approaches the speed of light at a rate that makes $\gamma$ diverge less rapidly than linearly in $t_r$. To be specific, we choose to work with
 \begin{equation}\label{deltax}
 \delta x(t_r)=-\frac{\epsilon}{b}\ln(\cosh(bt_r))~,
 \end{equation}
 where $\epsilon$ (controlling the amplitude of the perturbation) and $b$ (defining its characteristic time scale) are adjustable parameters. This shifts the quark/endpoint velocity by an amount $\delta v(t_r)=-\eps\tanh(bt_r)$, thus ensuring that in the remote past the net velocity asymptotes to $-1+\epsilon$ instead of $-1$.

 The exact, \emph{nonlinearized} string embedding (\ref{mikhsolnoncovariant}) dual to the deformed quark trajectory is
 \begin{eqnarray}\label{mikhaccelpert}
t(t_r,z)&=&t_r+\frac{z}{\sqrt{1
-\left(\frac{t_r}{\sqrt{A^{-2}+t_r^2}}-\eps\tanh(bt_r)\right)^2}}~,\\
X(t_r,z)&=&\sqrt{A^{-2}+t_r^2}-\frac{\epsilon}{b}\ln(\cosh(bt_r))+
\frac{\left(\frac{t_r}{\sqrt{A^{-2}+t_r^2}}-\epsilon\tanh(bt_r)\right)z}{\sqrt{1
-\left(\frac{t_r}{\sqrt{A^{-2}+t_r^2}}-\eps\tanh(bt_r)\right)^2}}~.\nonumber
\end{eqnarray}
{}From this, it is easy to see that in the region defined by the 3 conditions $t_r<0$, $-At_r\gg 1$ (so that we are almost on the asymptote of the hyperbolic trajectory (\ref{quarkaccel})) and $-bt_r\gg 1$ (so that the shift in the velocity is essentially time-independent), the string at fixed $t$ is nearly vertical and moves with approximately constant speed,\footnote{This statement holds independently of the value of $\epsilon$, and of the relation between $A$ and $b$.} $X(t,z)=-(1-\epsilon)t+\cO(1/A^2 t_r^2)+\cO(1/bt_r)$. Additionally, upon solving for $z$ in the first relation in (\ref{mikhaccelpert}), we find that when $0<\epsilon<1$ this vertical segment of string extends as expected all the way down to $z\to\infty$, due to the quark/endpoint behavior in the region $-At_r\gg 1/\sqrt{\epsilon}$.

\begin{figure}[htb]
\centering
\begin{tabular}{cc}
\includegraphics*[trim= 10 60 0 75,width=5.6cm]{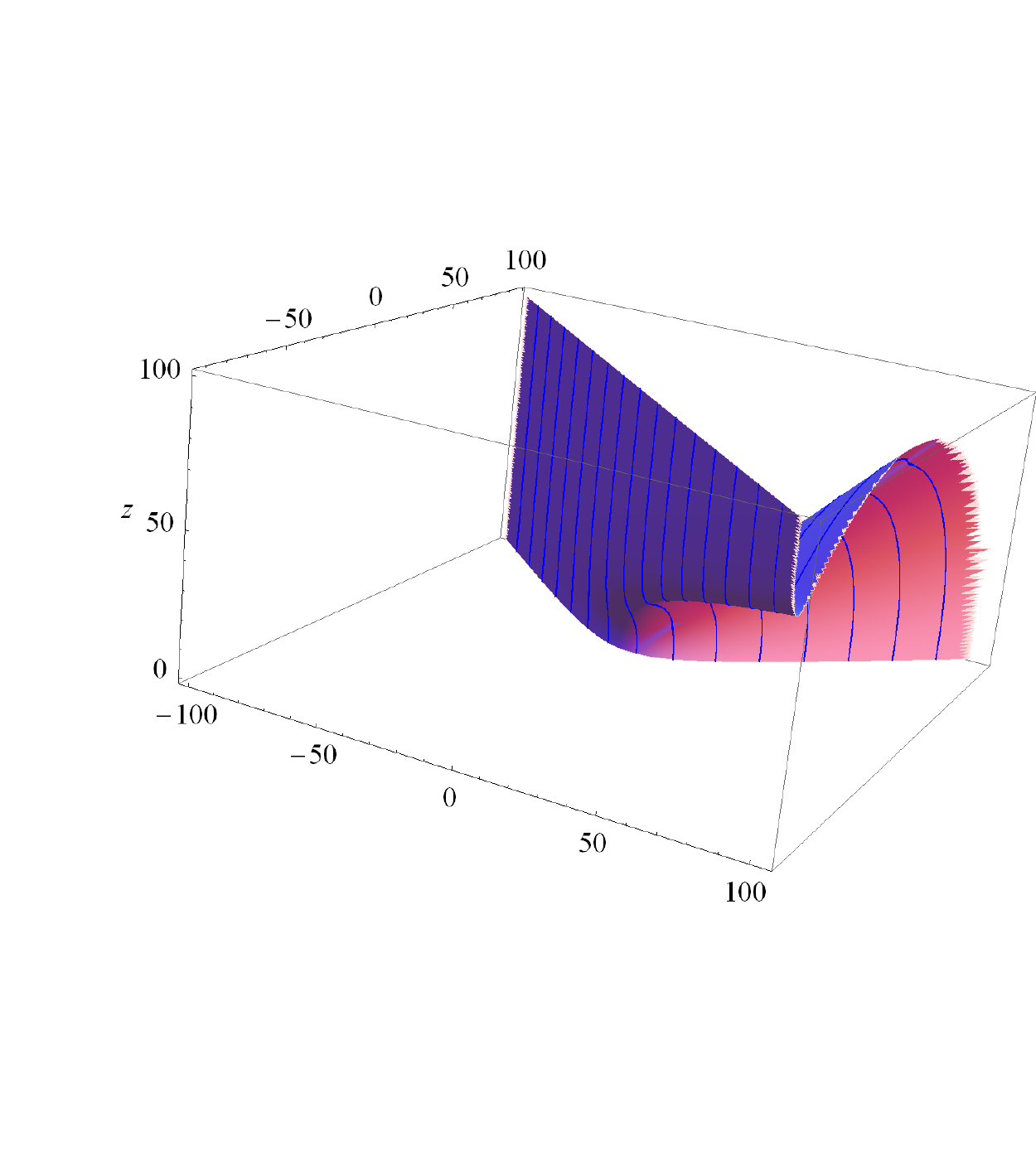}
&
\hspace*{0.5cm}
\includegraphics*[trim= 10 60 10 70,width=5.6cm]{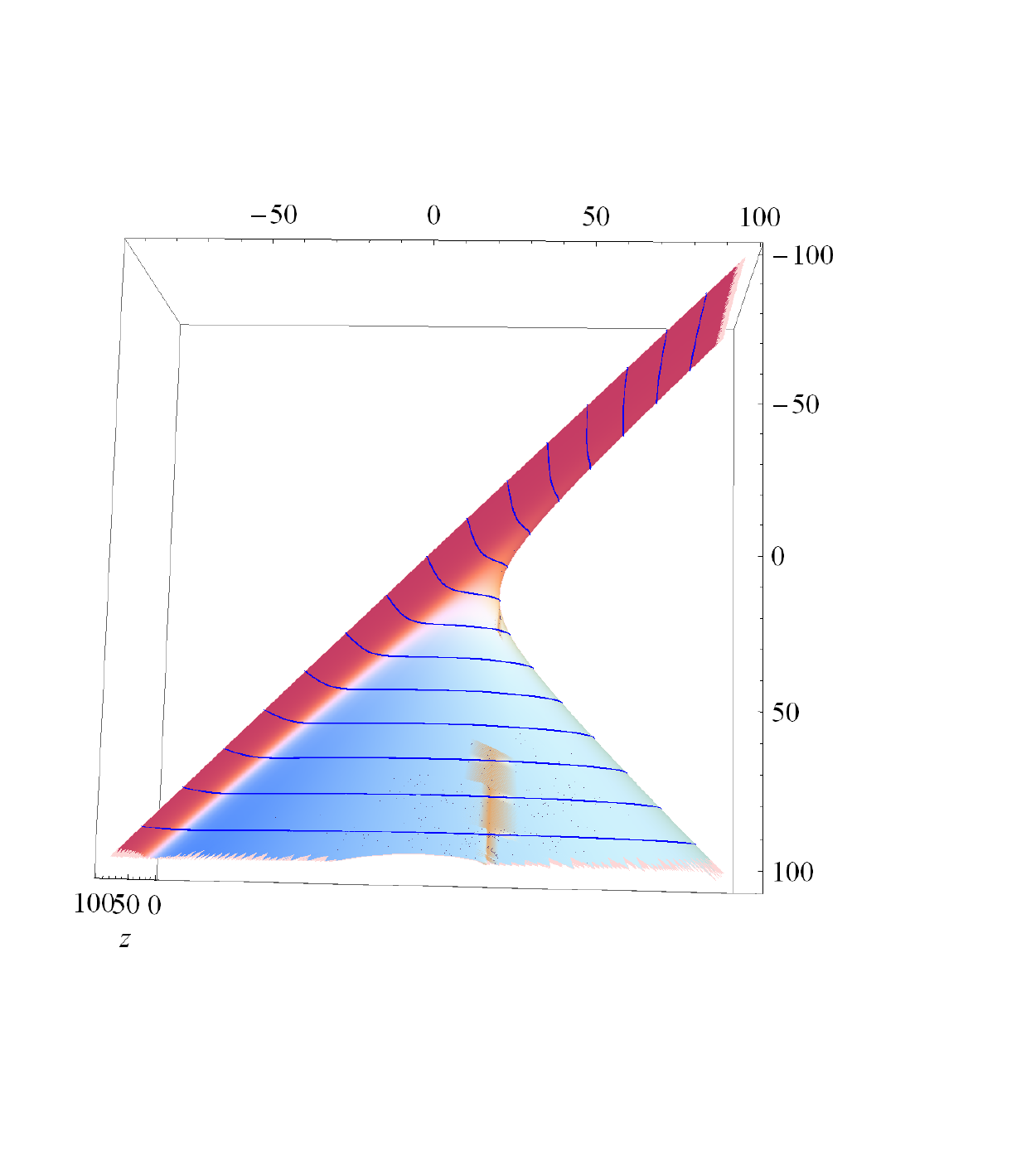}
\\
\includegraphics*[trim= 10 60 0 75, width=5.6cm]{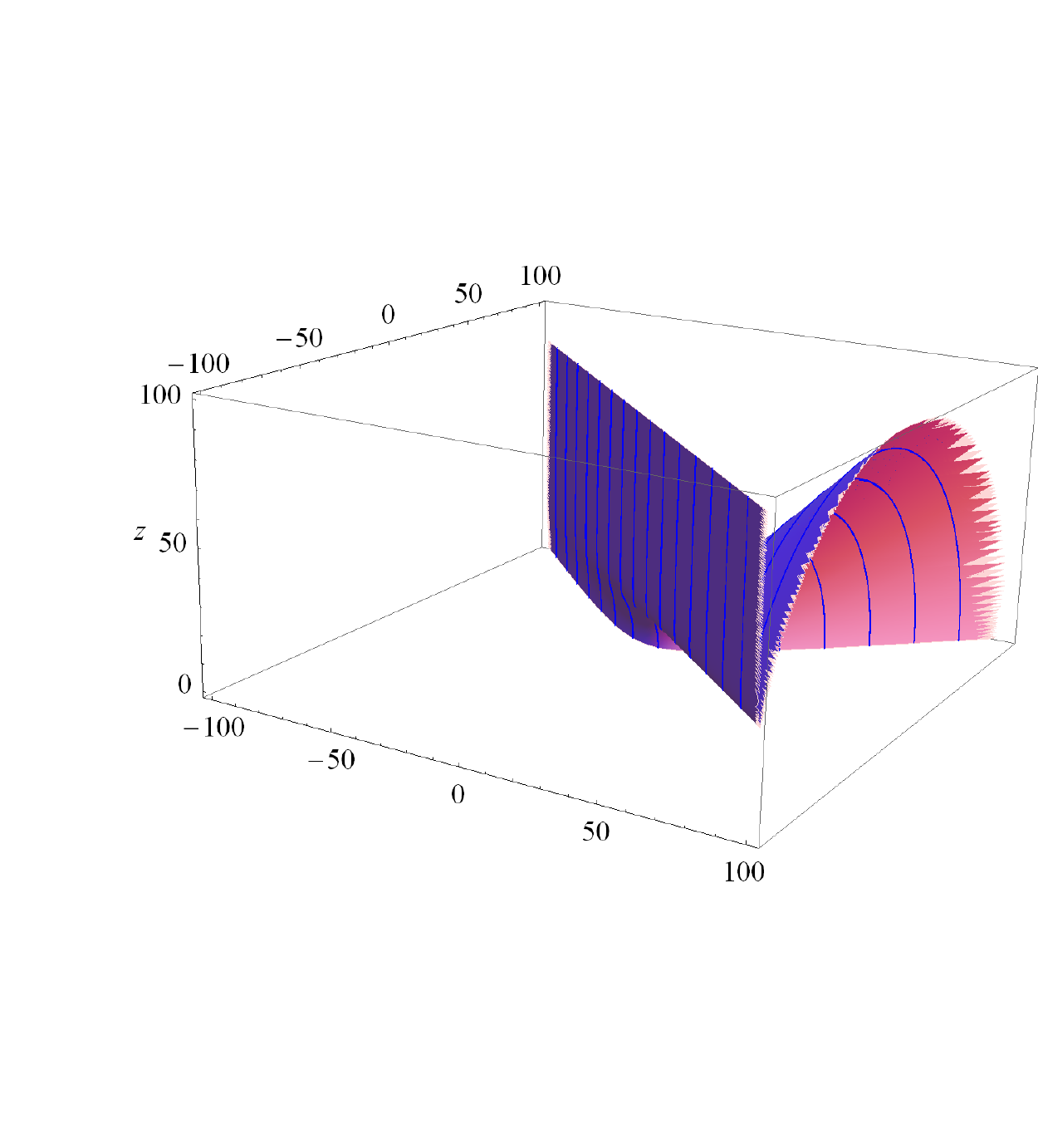}
&
\hspace*{-1cm}
\includegraphics*[trim= 10 60 10 62, width=5.6cm]{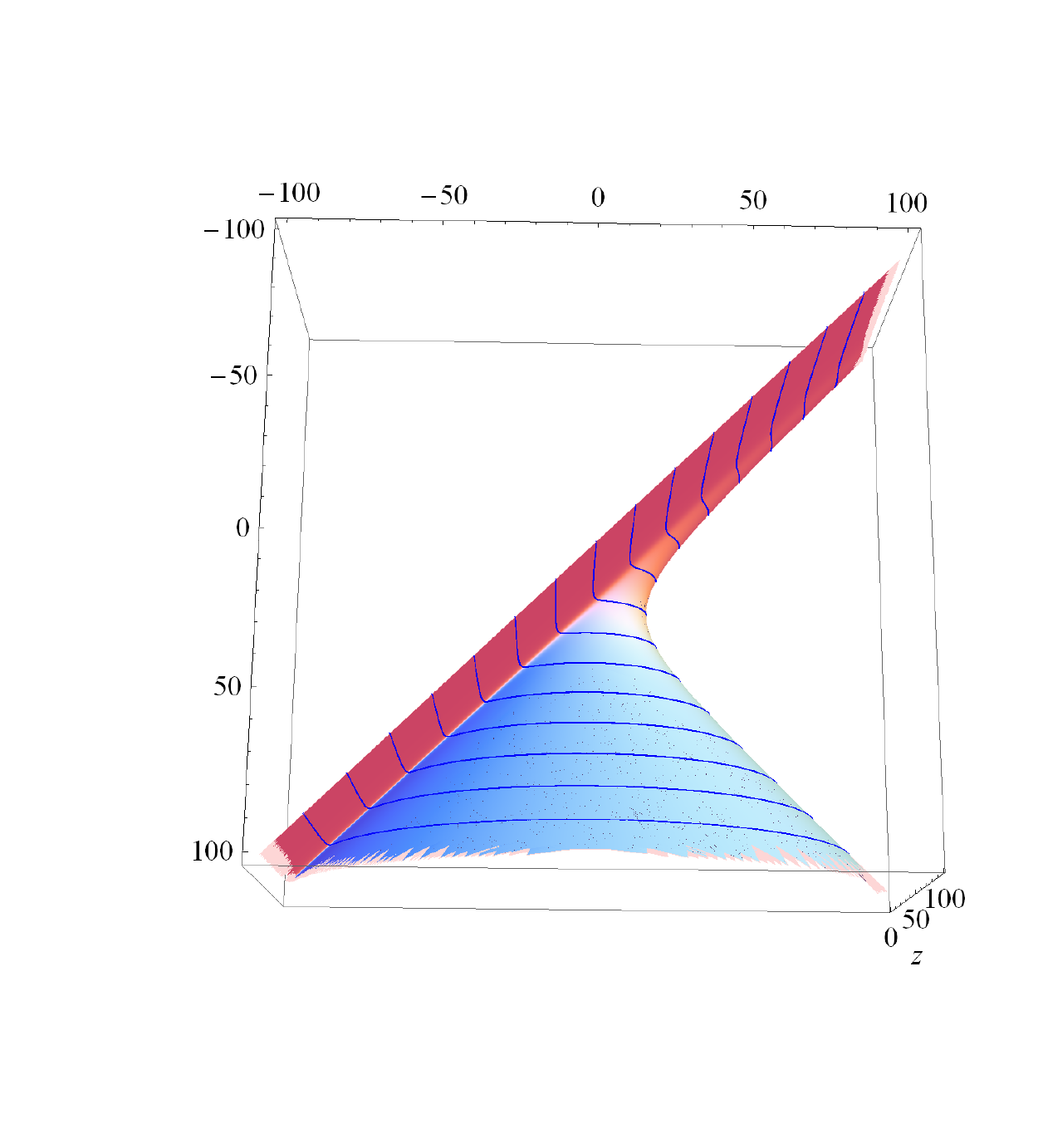}
\\
\includegraphics*[trim= 10 60 0 75, width=5.6cm]{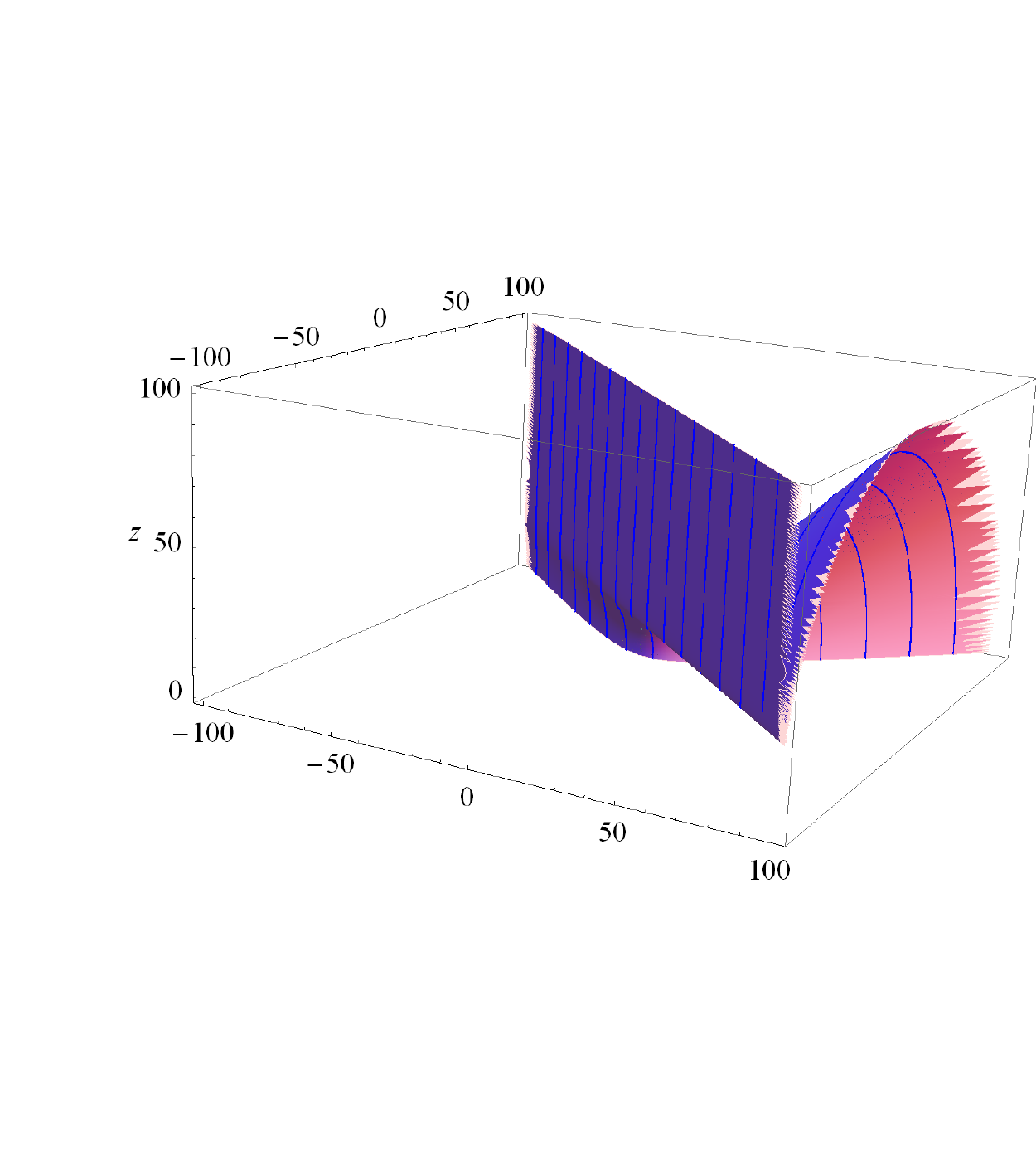}
&
\hspace*{1cm}
\includegraphics*[trim= 10 55 10 70, width=5.6cm]{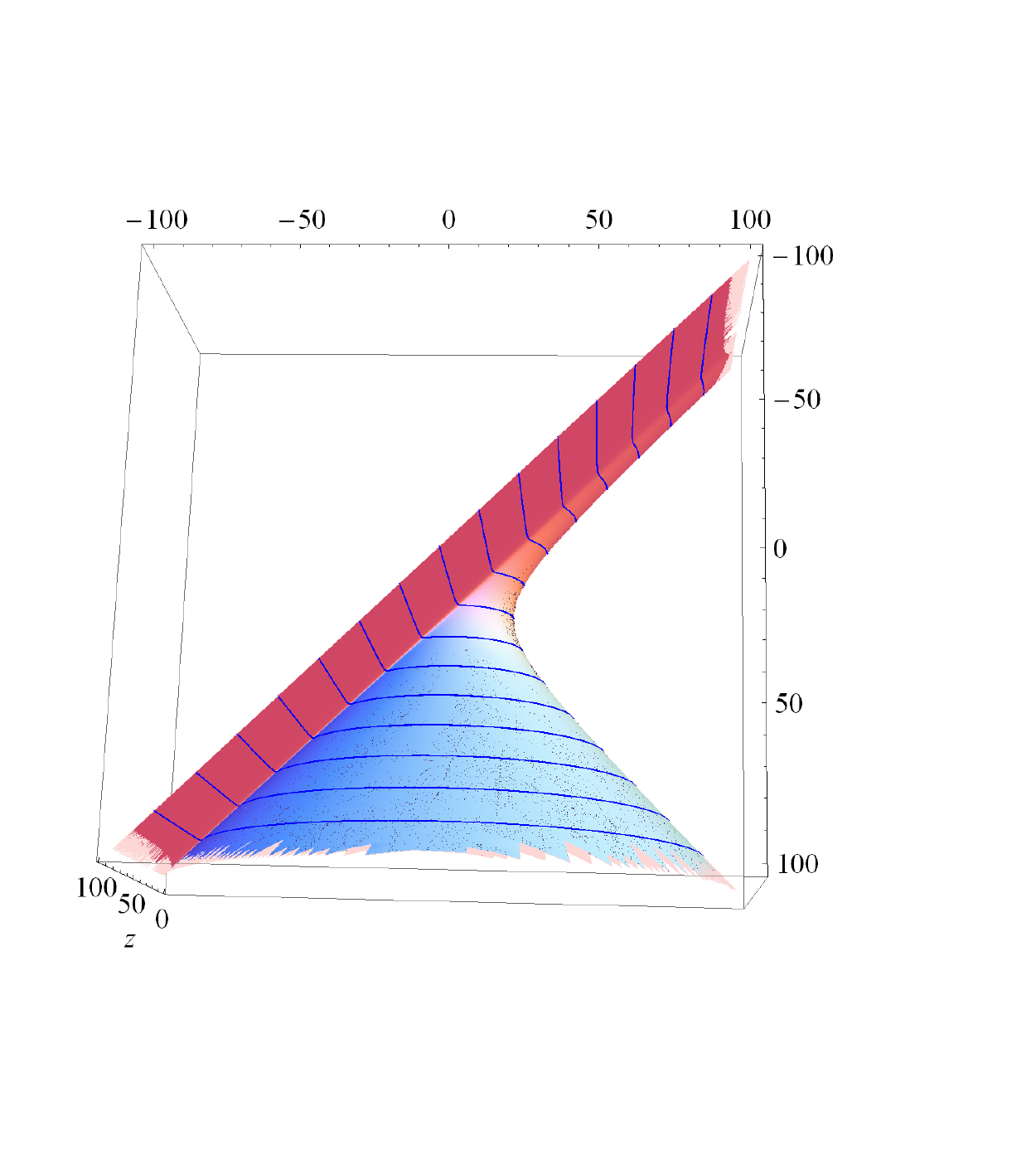}
\end{tabular}
\vspace*{-0.5cm}
\caption{Side and top views of the worldsheet (\ref{mikhaccelpert}) dual to a uniformly accelerated quark perturbed by (\ref{deltax}), in units such that $A=0.05$. The (blue) curves indicate string contours at fixed $t$, and the $z$ axis points upward. In all cases $b=1$, and, starting from the top, the rows correspond to $\epsilon=0.05,0.001,0.0001$. As the unperturbed embedding is approached, there remains a vertical wall extending upward from the horizon ($z_h=20$), which was missing in Fig.~\ref{uniformfig}.
}
\label{epsfig}
\end{figure}

For arbitrarily small $\epsilon$, this confirms that the string is continued by a vertical segment moving at constant velocity. As $\epsilon\to 0$, the juncture between this continuation and the portion of the string connected to the boundary approaches $z=A^{-1}$, and the range of retarded times which trace out the vertical segment recedes to $t_r\to-\infty$, which explains why this continuation is missing from the unperturbed embedding (\ref{mikhaccel}). All of the preceding statements can also be confirmed visually in the Mathematica plots shown in Fig.~\ref{epsfig}.

The  conclusion is that the situation is indeed as depicted in Fig.~\ref{arcscuspfig}.
The string is initially completely vertical and moving at the speed of light, with the segment at $z>A^{-1}$ describing the outer portion of the infinitely boosted Coulombic field that is taken as an initial condition at $t_r\to-\infty$ (having been set up by the behavior of the quark/endpoint \emph{prior to} the beginning of time). As the quark decelerates, there is a perturbation running downward along the string, and outward on the dual gluonic field configuration, but this perturbation never manages to catch up with the $z>A^{-1}$ segment, which therefore continues undisturbed.\footnote{The fact that the disturbance on the string remains above the horizon $z=z_h$ implies that in the case where the quark has a finite mass, where the string (rather than reaching the boundary) extends only up to the radial location $z_m$ given by (\ref{zm}), we must necessarily have $z_m\le z_h$. This might seem to predict a lower bound on the quark mass for a given acceleration $A$, but closer inspection shows that this is not so: for $z_m>0$, the horizon is at $z_h=\sqrt{A^{-1}+z_m^2}$ \cite{brownian}, so the condition $z_m\le z_h$ is always satisfied, independently of the value of the quark mass.}
 This is why a cusp develops. In short, \emph{the vertical string segment at $z>A^{-1}$ codifies a shock wave that loses contact with and is consequently shed by the quark.}

Let us be more precise on this last point. While it is evident that the vertical string segment that completes the string, extending from $z=A^{-1}$ and moving at the speed of light (according to $X(t,z)=-t$), describes the IR portion of a shock wave in the gluonic field, the total gluonic field at any given point is obtained by combining this contribution with that of the portion of arc traced out by the Mikhailov solution (\ref{mikhaccel}). At the level of one-point functions, this case is covered by the calculations carried out in \cite{iancu2,trfsq} for an arbitrary quark trajectory, using the embedding (\ref{mikhsol}). As noted in Section \ref{whippingsec}, it was found there that the entire result arises from a surface term that ends up being evaluated only at $z=0$. The authors of those works did not take into account, however, that for the special case of uniform acceleration the Mikhailov embedding is incomplete, so, when using (\ref{mikhaccel}) by itself, there would also be a surface term arising from $z=A^{-1}$.

This horizon contribution is thus missing from \cite{iancu2,trfsq}, but it is in fact precisely \emph{canceled} by the shock wave associated with the vertical completion of the embedding. This is clear from the fact that when our deformed embeddings (\ref{mikhaccelpert}) (that correctly extend all the way to $z\to\infty$) are used in \cite{iancu2,trfsq}, there is no net contribution from any intermediate point on the string, so as we take $\epsilon\to 0$ to approach the case of uniform acceleration, the same should be true by continuity. It can also be verified by directly evaluating the results of \cite{trfsq,iancu2} for the case of uniform acceleration, and noting that at asymptotic early times, $t\to-\infty$, the expressions given there correctly describe the \emph{full} boosted Coulombic profile associated with the quark moving at constant speed $v\to 1$, which is obviously dual to the \emph{complete} vertical string. The point then is that for uniform acceleration, the procedure of \cite{iancu2,trfsq} omits 2 features, which, when properly taken into account, cancel each other out.

It might seem peculiar that the vertical segment of the string, which in the AdS description very noticeably moves away from the quark/endpoint as $t\to\infty$, is in the end invisible in the gluonic field profiles determined in \cite{iancu2,trfsq}. This, however, is just a concrete illustration of the fact that local observables are \emph{not} sensitive to the instantaneous profile of the string, due to the delay associated with field propagation towards the AdS boundary. This implies that the direct interpretation of the complete string as a quark that sheds a shock wave, which is so naturally suggested by Fig.~\ref{arcscuspfig}, can really only be made in terms of an appropriate set of nonlocal probes of the gluonic field \cite{precursors}, whose importance we had already emphasized in Section \ref{whippingsec}. Related observations have been made very recently in \cite{iancu3}.

We should perhaps stress that there is of course nothing physically wrong with assuming as in \cite{xiao} that at $t\to-\infty$, in addition to a full vertical string $X(t,z)=-t$ corresponding to a quark traveling in the $-x$ direction at the speed of light, one has a second vertical string $X(t,z)=t$, dual to an \emph{antiquark} traveling in the $+x$ direction at light speed. If both string endpoints undergo uniform acceleration, then the complete string embedding will trace out the full semicircle (\ref{xiaoaccel}). Our point is just that this different initial condition \emph{does not} describe an isolated quark undergoing uniform acceleration.

Before closing this section, we should note that, because the termination of the embedding (\ref{mikhaccel}) depends solely on the quark behavior in the remote past, the main statements we have made here apply not just to the case of uniform acceleration, but also to any motion whatsoever that asymptotically approaches uniform acceleration as $t_r\to-\infty$ (and therefore has a speed $v\to 1$ at a rate such that $\gamma\propto -t_r$). For all such trajectories, (\ref{mikhsolnoncovariant}) will yield an incomplete (generally noncircular) embedding, which ought to be continued by a vertical string segment traveling at the speed of light (although generally the termination line $z=A^{-1}$ corresponding to $t_r\to-\infty$ will \emph{not} coincide with the worldsheet horizon).

 \section{Energy-Momentum Flow on the Worldsheet}\label{flowsec}

\begin{figure}[htb]
\centering
\includegraphics[width=12cm]{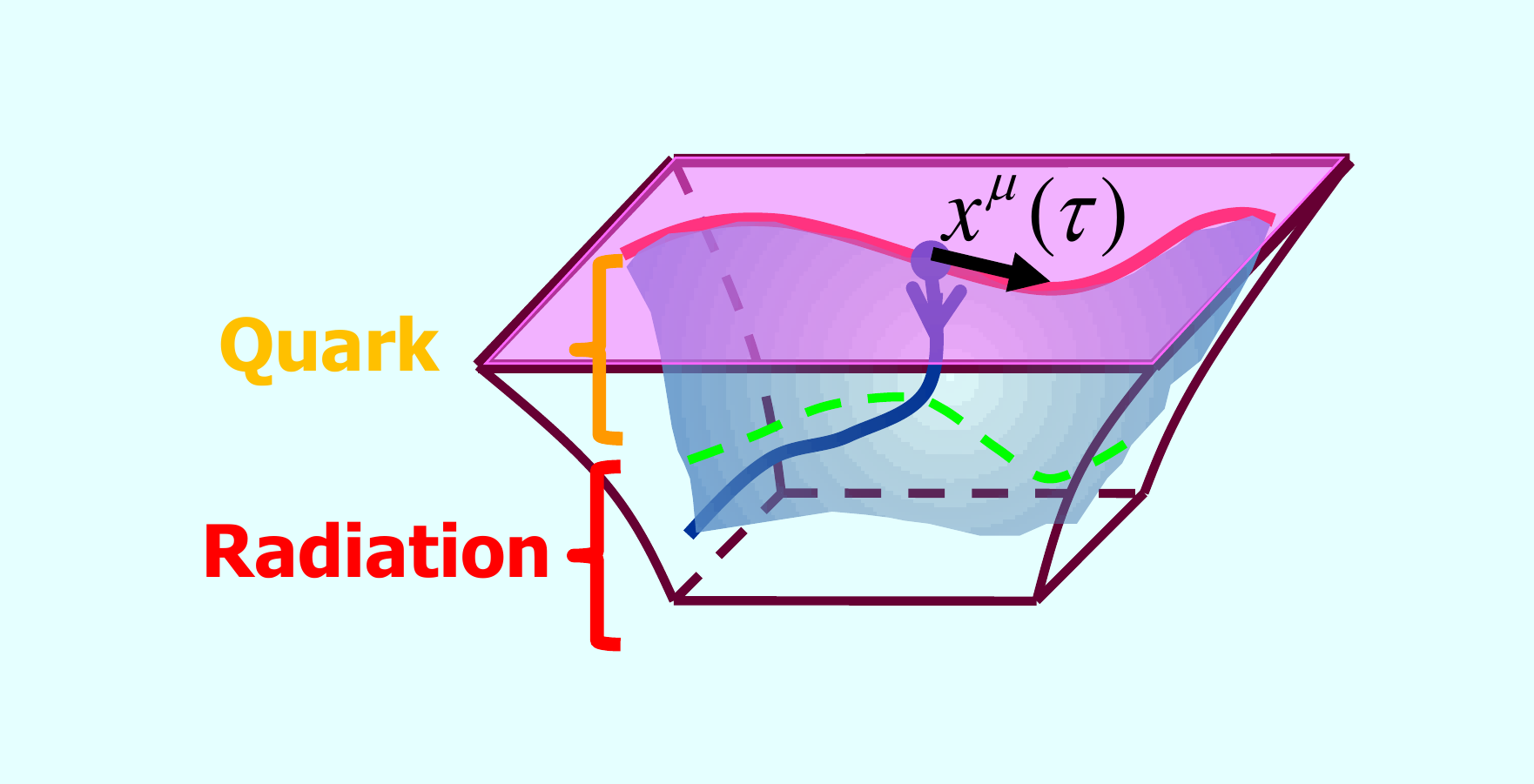}
 \caption{As explained in the main text, it is natural to try to interpret the portion of the string outside the worldsheet horizon (green dashed curve) as the quark, and the portion inside it as the radiation already emitted by the quark.
}
\label{horizonfig}
\end{figure}

As mentioned in the Introduction, whenever the quark accelerates, a horizon develops on the worldsheet of the dual string \cite{dragtime}, and this horizon has often been regarded as the curve that separates the portion of the string that corresponds to the quark ($+$ near field) proper from the portion that corresponds to the radiated field \cite{dragtime,xiao,beuf,veronikamukund,iancu2,rajagopalshining}. This is indicated schematically in Fig.~\ref{horizonfig}. In support of this interpretation, it was shown in \cite{xiao} that, for the particular case of uniform acceleration, the rate at which energy crosses the horizon agrees with the quark radiation rate known from the work of \cite{mikhailov,dragtime}, i.e., the Lienard formula (\ref{lienard}). It is thus natural to wonder whether the same is true, for the horizon or for some other curve, when we consider more general quark/endpoint motions.

Before proceeding, it is useful to explicitly state  the minimal criteria that would entitle us to identify a given curve on the string worldsheet as a dividing line of the sort described above. In an arbitrary parametrization, we will denote the dividing curve as $\sigma_{\scriptstyle \mathrm{d}}(\tau)$. For concreteness, let us assume that $\sigma=0$ denotes the string endpoint at the AdS boundary ($z=0$).  This curve should be such that the string segment \emph{outside} of it, $\sigma<\sigma_{\scriptstyle \mathrm{d}}(\tau)$ (i.e., the piece of the string closer to the AdS boundary, corresponding to the UV region of the gauge theory), can be directly interpreted as dual to the quark; whereas the segment \emph{inside} of it, $\sigma>\sigma_{\scriptstyle \mathrm{d}}(\tau)$ (extending all the way to $z\to\infty$, corresponding to the IR region of the gauge theory), encodes all of the gluonic radiation previously emitted by the quark.

Knowing that the density and rate of flow of the spacetime energy-momentum are given by the canonical momentum densities $\Pi^{\tau}_{\mu}$ and $\Pi^{\sigma}_{\mu}$ following from (\ref{nambugoto}), we thus require that $\sigma_{\scriptstyle \mathrm{d}}(\tau)$ satisfy
\begin{eqnarray}\label{dividing}
\int_0^{\sigma_{\scriptscriptstyle \mathrm{d}}(\tau)} \!\!\!\! d\sigma\,\Pi^{\tau}_{\mu}(\tau,\sigma)& = &mv_{\mu}(\tau_q)~, \\
\left(\Pi^{\sigma}_{\mu}-\Pi^{\tau}_{\mu}\frac{d\sigma_{\scriptstyle \mathrm{d}}}{d\tau}\right)\Big|_{\sigma_{\scriptscriptstyle \mathrm{d}}(\tau)}& = & \frac{\sqrt{\lambda}}{2\pi}a^2(\tau_q) v_{\mu}(\tau_q)\frac{d\tau_q}{d\tau}~.\nonumber
\end{eqnarray}
The first of these conditions demands that the total 4-momentum of the exterior string segment at time $\tau$ correctly reproduce the quark 4-momentum (with a divergent quark mass, following from (\ref{zm}) with $z_m\to 0$). The second condition requires the net rate of 4-momentum flowing across the dividing curve (taking into account both the intrinsic flow along the string at that location, and the 4-momentum density swept by the motion of the curve itself,) to equal the Lienard rate (\ref{lienard}).

At first sight, it  seems natural to evaluate the 4-velocity and 4-acceleration of the quark appearing in the right-hand side of (\ref{dividing}) at $\tau_q=\tau$, meaning that they match onto the quark 4-momentum and radiation rate at the same instant $\tau$ when the profile of the string is being examined. However, we know from Section \ref{mikhailovsec} that quark/endpoint information propagates along the fixed $\tau_r$ (Mikhailov) lines on the worldsheet, so the point on the string at $(\tau,\sigma_{\scriptstyle \mathrm{d}}(\tau))$ only knows what the quark/endpoint was doing at the corresponding retarded time $\tau_r$. The canonical momentum densities depend only on this local information. So, unless the curve $\sigma_{\scriptstyle \mathrm{d}}(\tau)$ itself somehow contains \emph{global} information about the quark worldline, then, by causality, we can at best hope to enforce (\ref{dividing}) with the quantities pertaining to the quark evaluated at $\tau_q=\tau_r$. The derivative $d\tau_q/d\tau$ in the second condition is then included to convert the rate per unit $\tau_q$ given by the Lienard formula to a rate per unit $\tau$ (which is what the left-hand side yields).

We can now establish, in any given parametrization, whether these 2 conditions are satisfied for some candidate curve, such as the worldsheet horizon. It is natural to start by considering the covariant gauge in which the retarded Mikhailov solution (\ref{mikhsol}) is constructed: $\tau=\tau_r$, $\sigma=z$. The horizon is found by noting from (\ref{wsmetric}) that null curves on the worldsheet satisfy $(z^2 a^2-1)d\tau_r^2-2dzd\tau_r=0$. One class of solutions has $d\tau_r=0$, describing the Mikhailov lines, which are tangent to the \emph{inward}-pointing lightcones, i.e., those that point towards increasing $z$. The worldsheet horizon $z_h(\tau_r)$ is among the other class of solutions, describing the \emph{outward}-pointing lightcones, i.e., those that ordinarily point towards the AdS boundary. It therefore satisfies the differential equation
\begin{equation}\label{wshorizon}
\frac{dz_h}{d\tau_r}={1\over 2}(z_h^2 a^2-1)~,
\end{equation}
with final condition chosen to reflect the asymptotic region on the string to which fluctuations can escape. E.g., if the quark is accelerated for some period of time and subsequently let free for all $\tau_r>\tau_{\scriptstyle \mathrm{release}}$, then we must have $z_h\to\infty$ at $\tau_r=\tau_{\scriptstyle \mathrm{release}}$ \cite{dragtime}. For the specific case of constant acceleration, $a^2=A^{2}$, equation (\ref{wshorizon}) is solved by $z_h=1/A$, confirming the presence of a horizon at this radial location, just as was stated in the previous section.

In these coordinates, the canonical momentum densities associated to (\ref{nambugoto}) read
\begin{eqnarray}\label{pisgeneral}
\Pi^{\tau_r}_{\mu}& \equiv &\frac{\p\mathcal{L}}{\p(\p_{\tau_r}X^{\mu})}=
\frac{\sqrt{\lambda}}{2\pi}\frac{\left(1+(\p_z X)^2\right)\p_{\tau_r}X_{\mu}
-(\p_{\tau_r}X\cdot\p_{z}X)\p_z X_{\mu}}{z^2\sqrt{(\p_{\tau_r}X\cdot\p_{z}X)^2
-(\p_{\tau_r} X)^2\left(1+(\p_z X)^2\right)}}~, \\
\Pi^{z}_{\mu}& \equiv &\frac{\p\mathcal{L}}{\p(\p_{\tau_r}X^{\mu})}=
\frac{\sqrt{\lambda}}{2\pi}\frac{(\p_{\tau_r} X)^2\p_{z}X_{\mu}
-(\p_{\tau_r}X\cdot\p_{z}X)\p_{\tau_r} X_{\mu}}{z^2\sqrt{(\p_{\tau_r}X\cdot\p_{z}X)^2
-(\p_{\tau_r} X)^2\left(1+(\p_z X)^2\right)}}~.\nonumber
\end{eqnarray}
Using the explicit form of (\ref{mikhsol}), they simplify to
\begin{eqnarray}\label{pismikh}
\Pi^{\tau_r}_{\mu}& = &
\frac{\sqrt{\lambda}}{2\pi z^{2}}v_{\mu}~, \\
\Pi^{z}_{\mu}&=&\frac{\sqrt{\lambda}}{2\pi}\left(\frac{a_{\mu}}{z}+a^2 v_{\mu}\right)~,\nonumber
\end{eqnarray}
with the expressions on the right-hand side evaluated as usual at the appropriate retarded time $\tau_r$.
{}From the first expression in (\ref{pismikh}), and using (\ref{zm}) with $z_m\to 0$, we see that when we consider a snapshot of the string at a given value of $\tau_r$ (i.e., a Mikhailov line), the total 4-momentum of the portion of the string outside $z_d(\tau_r)$ is $mv_{\mu}-\sqrt{\lambda}/2\pi z_d$, so our first condition in (\ref{dividing}) is  satisfied only if $z_d\to\infty$. Interestingly, with this choice, the expression for $\Pi^{z}_{\mu}$ in (\ref{pismikh}) directly reduces to the Lienard formula, so our second condition in (\ref{dividing}) is also satisfied. We thus learn that, \emph{when we slice the worldsheet at fixed $\tau_r$, the entire string is dual to the quark ($+$ near field), and the radiation is located at the Poincar\'e horizon $z\to\infty$}.\footnote{We refrain from saying that the radiation lies in the region of global AdS beyond the Poincar\'e wedge because, strictly speaking, the CFT on Minkowski can only describe the accumulation of energy-momentum at $z\to\infty$ as seen by a Poincar\'e observer, and not the actual crossing.} This means in particular that, in the Mikhailov parametrization, the dividing line does \emph{not} generically coincide with the horizon (\ref{wshorizon}).

Notice that, according to (\ref{mikhsolnoncovariant}), going to $z\to\infty$ at fixed $\tau_r$ implies observing the gauge theory at time $t\to\infty$, which is why in the $(\tau_r,z)$ gauge the radiation is found to lie at the extreme IR. The conjecture that the horizon plays the role of dividing curve would thus have more of a fighting chance in the static gauge choice $\tau=t$, $\sigma=z$, where the worldsheet is parametrized with coordinates that are more directly relevant to the gauge theory. The work \cite{xiao} concerned itself with the rate of energy flow per unit CFT observation time $t$, and not per unit retarded quark proper time $\tau_r$. And even the separation (\ref{emikh}) achieved in \cite{mikhailov,dragtime} between intrinsic and radiated energy followed from an analysis of the string at fixed $t$, rather than fixed $\tau_r$.

To change the worldsheet parametrization from $(\tau=\tau_r, \sigma=z)$ to $(\bar{\tau}=t, \bar{\sigma}=z)$, we first deduce from (\ref{mikhsolnoncovariant}) and $d\tau_r=dt_r/\gamma$ that
\begin{eqnarray}\label{coordchange}
\left(\frac{\p\tau_r}{\p t}\right)_z&=&
   \frac{1}{\gamma+\gamma^4\vec{v}\cdot\vec{a}z}~, \qquad
   \left(\frac{\p\tau_r}{\p z}\right)_t=
      -\frac{1}{1+\gamma^3\vec{v}\cdot\vec{a}z}~,\\
\left(\frac{\p z}{\p t}\right)_z&=&0~, \qquad\qquad\qquad\quad
    \left(\frac{\p z}{\p z}\right)_t\,=1~,\nonumber
\end{eqnarray}
and then use this in the transformation law for the momentum densities (see, e.g., \cite{dragqqbar}) to infer that\footnote{To avoid possible misunderstandings, we should perhaps stress that in these expressions $\vec{a}$ refers to the 3-acceleration, and is therefore distinct from the spatial components of the 4-acceleration $a_{\mu}=\gamma^4(-\vec{v}\cdot\vec{a},
\vec{v}\cdot\vec{a}\,\vec{v}+\gamma^{-2}\vec{a}\,)$, whose squared norm is denoted by $a^2\equiv a_{\mu}a^{\mu}$. Likewise, $\vec{v}$ differs from the spatial components of $v_{\mu}=\gamma(-1,\vec{v}\,)$.}
\begin{eqnarray}\label{pibart}
\bar{\Pi}^t_{\mu}&=&\left(\frac{\p z}{\p z}\right)_t \Pi^{\tau_r}_{\mu}
-\left(\frac{\p\tau_r}{\p z}\right)_t \Pi^z_{\mu}\nonumber\\
&=& \frac{\sqrt{\lambda}}{2\pi} \left[\frac{v_{\mu}}{z^2}
    +\frac{1}{1+\gamma^3\vec{v}\cdot\vec{a}z}
    \left(\frac{a_{\mu}}{z}+a^2 v_{\mu}\right)\right]
\end{eqnarray}
and
\begin{eqnarray}\label{pibarz}
\bar{\Pi}^z_{\mu}&=&\left(\frac{\p \tau_r}{\p t}\right)_z \Pi^z_{\mu}
-\left(\frac{\p z}{\p t}\right)_t \Pi^{\tau_r}_{\mu}\nonumber\\
&=& \frac{\sqrt{\lambda}}{2\pi} \frac{1}{\gamma+\gamma^4\vec{v}\cdot\vec{a}z}
    \left(\frac{a_{\mu}}{z}+a^2 v_{\mu}\right)~.
\end{eqnarray}

For uniform acceleration, it is easy to check that the rate of energy flow per unit $t$ at the worldsheet horizon $z_h=A^{-1}$ following from (\ref{pibarz}) indeed agrees with the result of \cite{xiao},
\begin{equation}\label{piztxiao}
\bar{\Pi}^z_t|_{z=A^{-1}}=-\frac{\sqrt{\lambda}}{2\pi}A^2~,
\end{equation}
which is essentially the (non-covariant) Lienard result, except for 2 features that we will point out shortly. On the other hand, the form of (\ref{pibarz}) presages no such agreement at the horizon (\ref{wshorizon}) associated with a general quark/endpoint motion, although at this point it is difficult to make a definite statement to this effect, given that the horizon is
determined by the global causal structure of the worldsheet, rather than the local behavior of the quark/endpoint at a given retarded time.

To make progress, let us start from the observation that what is special about the case of uniform acceleration is that the worldsheet geometry is static, and because of this, the horizon coincides with the stationary-limit curve $z_{\scriptstyle \mathrm{ergo}}(\tau_r)$ defined by the condition $g_{\tau_r\tau_r}=0$, beyond which it is $z$ and not $\tau_r$ that plays the role of timelike coordinate. {}From (\ref{wsmetric}) we see that this curve, where the outward lightcone becomes horizontal, is located at
\begin{equation}\label{slc}
z_{\scriptstyle \mathrm{ergo}}=\frac{1}{\sqrt{a^2}}~,
\end{equation}
which shows that, unlike the horizon, it is determined by purely local endpoint/quark information.
For $a\neq 0$, (\ref{slc}) coincides with the horizon (\ref{wshorizon}) only if $a\cdot j=0$, or equivalently, if $dz_h/d\tau_r=0$, which is what happens for uniform acceleration. Generally there is a portion of the stationary-limit curve outside the horizon, $z_{\scriptstyle \mathrm{ergo}}<z_h$, and the region between these 2 curves is the one-dimensional analog of an ergosphere \cite{dragtime} (a concept whose relevance was first noted, in the $T > 0$ AdS/CFT context, in \cite{argyres2}).

After some algebra, the general expression for the rate of energy flow (\ref{pibarz}) at $z_{\scriptstyle \mathrm{ergo}}$ can be seen to take the form
\begin{equation}\label{pislc}
\bar{\Pi}^z_t|_{z=1/\sqrt{a^2}}=-\frac{\sqrt{\lambda}}{2\pi}a^2~.
\end{equation}
This is evidently the generalization of (\ref{piztxiao}) to the case of arbitrary motion. We thus learn that, for generic quark/endpoint motion, the contact established in \cite{xiao} between energy flow (per unit $t$) on the string and the Lienard rate of energy emission takes place not at the horizon, but at the stationary-limit curve on the string worldsheet.\footnote{Given the way in which the analysis of \cite{dragtime,lorentzdirac,damping} uses auxiliary data at the AdS boundary, it follows automatically that $\bar{\Pi}^z_t$ at the stationary-limit curve also reproduces the energy flow given by the modified Lienard formula relevant at finite quark mass.}

Notice that the right-hand side of (\ref{pislc}) is evaluated at the retarded time $t_r$ (or $\tau_r$) corresponding through (\ref{mikhsolnoncovariant}) to the given values of $t$ and $z$. This means that, when we consider the string at a time $t$ and determine the instantaneous rate of energy through the point of the string located on the stationary-limit curve, the result reflects the rate of energy radiated by the quark not at $t$, but at the earlier time $t_r$.
As we had already noted above, by causality, it could not have been any other way: information on the worldsheet propagates along Mikhailov lines, so the behavior at any given point $(t,z)$ is determined by the quark/endpoint behavior at the earlier time $t_r$ obtained by following the corresponding Mikhailov line all the way back to the AdS boundary. No such consideration was necessary in \cite{xiao}, but that was only because the rate (\ref{piztxiao}) is independent of time.

The form of (\ref{pislc}) seems to raise the hope that $z_{\scriptstyle \mathrm{ergo}}$ might be the dividing curve we are looking for, but unfortunately, under closer scrutiny this hope quickly evaporates, for several reasons. First, despite appearances, the expression on the right-hand side of (\ref{pislc}) does not quite agree with the Lienard formula (\ref{lienard}). It would give the rate of energy radiated per unit retarded time $t_r$, but the left-hand side is by definition a rate per unit $t$. In other words, the formula is missing the factor of $dt_r/dt$ that corresponds to the $d\tau_q/d\tau$ in (\ref{dividing}). Second, (\ref{pislc}) does not incorporate the energy swept by the putative dividing line as it moves, and therefore does not meet our second condition in (\ref{dividing}). Third, for motions in which the quark approaches constant velocity at $t\to\pm\infty$, the stationary-limit curve $z_{\scriptstyle \mathrm{ergo}}$ generically cuts the string at 2 points (corresponding to 2 different retarded times), delimiting a finite region \cite{dragtime}, so one would need to modify (\ref{dividing}) and take into account the flow of energy at both edges. (The physical point here is that the string is vertical both close to and very far from the boundary, encoding a purely Coulombic profile, with no radiation.) Fourth, even if we ignore the previous 3 problems, the sign in (\ref{pislc}) comes out wrong: it indicates a flow of positive energy towards the region that is supposed to be identified with the quark, $z<z_{\scriptstyle \mathrm{ergo}}$, in spite of the fact that the actual quark is losing energy to radiation throughout its entire motion.
Fifth, unlike the energy, the rate of spatial momentum flow $\bar{\Pi}^z_i$ at $z_{\scriptstyle \mathrm{ergo}}$ does not turn out to even come close to the corresponding Lienard formula.\footnote{There is no contradiction with Lorentz covariance, because the canonical momentum density $\bar{\Pi}^z_{\mu}$ does not in fact transform as a dual vector under Lorentz transformations: there is an additional contribution due to the change in static gauge from $t$ to $t'$ (see, e.g., \cite{dragqqbar}).}

It is clear then that the stationary-limit curve cannot play the role of dividing line for arbitrary quark motion, so we must go back to the original conjecture that the horizon itself is the dividing curve. The fact that the horizon is determined by global information about the quark trajectory makes it difficult to locate its position, but also leaves the door open to the possibility that the horizon might conceivably satisfy our conditions (\ref{dividing}) evaluated at time $t$ instead of $t_r$, which is what seems to have been envisioned by most authors. The simplest way to test this is to examine the one example where we do know where the horizon is located: the case of uniform acceleration. Using (\ref{xiaoaccel}), we can derive the relevant densities directly in terms of $t$. The $\mu=0$ components are found to reduce to
\begin{eqnarray}\label{pibarsxiao}
\bar{\Pi}^t_t&=&-\frac{\sqrt{\lambda}}{2\pi}\frac{1+A^2 t^2}{z^2\sqrt{1+A^2 t^2 - A^2 z^2}}~,\\
\bar{\Pi}^z_t&=&\mp\frac{\sqrt{\lambda}}{2\pi}\frac{A^2 t}{z\sqrt{1+A^2 t^2 - A^2 z^2}}~.\nonumber
\end{eqnarray}
With this, it is easy to check that our first condition in (\ref{dividing}) is not satisfied, so the portion of string outside the horizon cannot be equated with the quark. Additionally, given that $z_{\scriptstyle \mathrm{ergo}}=z_h$ for this case, the fourth and fifth problems that we found above for the stationary-limit curve are present here as well.

 The counterexample of the previous paragraph is enough to disprove the conjecture that the horizon plays the role of dividing line on the string at fixed $t$. But we can point out another difficulty of this proposal for generic quark trajectories that asymptote to a constant velocity at $t\to-\infty$: because of its global nature, the horizon descends from $z\to\infty$ \emph{before} the quark starts to radiate \cite{dragtime}, contradicting the idea that the rate of 4-momentum flow through the horizon should represent the radiation emitted by the quark at the same time $t$. We could of course solve this problem by simply dropping the $dz_{\scriptstyle \mathrm{d}}/dt$ term from our second condition in (\ref{dividing}), but that term was actually the only reason why we were able to maintain a feeble hope that the horizon might work as a dividing line at fixed $t$. If it is a matter of looking at $\bar{\Pi}^z_t$ alone, then (\ref{pibarz}) and causality disprove the conjecture for arbitrary quark worldlines. As a matter of fact, no choice whatsoever of $z_{\scriptstyle \mathrm{d}}(t)$ can make $\bar{\Pi}^z_t$ match the required Lienard formula (\ref{lienard}) (evaluated at $t$ or $t_r$) for generic trajectories.

The overall conclusion of this section is thus that, \emph{when we examine the string at fixed $t$, the separation between the near and radiation fields of the quark is not reflected geometrically on the worldsheet (with a string segment corresponding directly to each contribution), and can be accessed only through the algebraic-differential procedure of \cite{mikhailov,dragtime}.}\footnote{An alternative is to carry out the explicit computations of the gluonic field profile, as in \cite{iancu2,trfsq,tmunu}. In this procedure, the data at some observation time $t_{\scriptstyle \mathrm{obs}}$ in the gauge theory are determined by adding up the contributions of points of the string at all previous times $t$, associated with even earlier endpoint times $t_r$.}

\section*{Acknowledgements}
We are grateful to Mariano Chernicoff and Veronika Hubeny for useful discussions. We also thank Mariano Chernicoff for comments on the manuscript.
The present work was partially supported by Mexico's National Council of Science and Technology (CONACyT) grant 104649, as well as DGAPA-UNAM grant IN110312.

\end{document}